\documentclass[unnumsec,webpdf,modern,medium,namedate]{oup-authoring-template}

\usepackage{amsmath}
\usepackage{amssymb}
\usepackage{amsfonts}
\usepackage{xcolor}
\usepackage{float}
\usepackage{booktabs}
\usepackage{graphicx}
\usepackage{subfigure}
\usepackage{color, soul}
\usepackage{tabularx}
\graphicspath{{Fig/}}

\theoremstyle{thmstyleone}%
\theoremstyle{thmstyletwo}%
\theoremstyle{thmstylethree}%

\begin{document}

\journaltitle{arXiv}
\copyrightyear{2024}
\pubyear{2024}
\appnotes{Paper}

\firstpage{1}


\title[Seismic Image Denoising With A Physics-Constrained Deep Image Prior]{Seismic Image Denoising With A Physics-Constrained Deep Image Prior}

\author[1,$\ast$]{Dimitri P. Voytan}
\author[2]{Sriram Ravula}
\author[3]{Alexandru Ardel}
\author[4]{Elad Liebman}
\author[5]{Arnab Dhara}
\author[6]{Mrinal K. Sen}
\author[7]{Alexandros Dimakis} 

\authormark{Voytan et al.}

\address[1,3,4]{\orgname{SparkCognition}, \orgaddress{\street{2708 Riata Vista Cir Suite B-100}, \postcode{78727}, \state{Austin, TX}, \country{United States of America}}}
\address[2,6,7]{\orgdiv{Cockrell School of Engineering}, \orgname{The University of Texas at Austin}, \orgaddress{\street{2501 Speedway, EER Building Room}, \postcode{78713-8924}, \state{Austin, TX}, \country{United States of America}}}
\address[1,5,6]{\orgdiv{University of Texas Institute for Geophysics \& John A. and Katherine G. Jackson School of Geosciences}, \orgname{The University of Texas at Austin}, \orgaddress{\street{University Station, Box X}, \postcode{78713-8924}, \state{Austin, TX}, \country{United States of America}}}

\corresp[$\ast$]{Corresponding author. \href{email:email-id.com}{dvoytan@sparkcognition.com}}




\abstract{Seismic images often contain both coherent and random artifacts which complicate their interpretation. To mitigate these artifacts, we introduce a novel unsupervised deep-learning method based on Deep Image Prior (DIP) which uses convolutional neural networks. Our approach optimizes the network weights to refine the migration velocity model, rather than the seismic image, effectively isolating meaningful image features from noise and artifacts. We apply this method to synthetic and real seismic data, demonstrating significant improvements over standard DIP techniques with minimal computational overhead.}

\keywords{Seismic Imaging, Denoising, Machine Learning, Unsupervised Deep Learning}

\maketitle

\section{Introduction}

Seismic imaging is routinely employed to characterize the geologic structure of the shallow subsurface. One widely used method is reverse-time migration (RTM), which is a wave-equation-based imaging method that excels in strongly heterogeneous media \cite[]{baysal1983reverse, loewenthal1983reversed, mcmechan1983migration, whitmore1983iterative, zhou2018reverse}. However, RTM images often have artifacts that complicate interpretation. These include long wavelength smearing from cross-correlation of the forward and time-reversed wavefields, short-wavelength parabolic swings,  acquisition footprints due to recording aperture limitations, and random noise. Classical mitigation strategies for long wavelength artifacts include muting high amplitude events at far offsets \cite[]{yoon2004challenges, yoon2006reverse}, modifying the imaging condition via wavefield decomposition or with Poynting-vector based angle weighting \cite[]{liu2011effective, yoon2006reverse}, and post-processing with a Laplacian filter \cite[]{youn2001depth, zhang2009practical}.

Least-squares migration (LSM) can address some of these issues, producing images with more balanced amplitudes and mitigated acquisition footprint \cite[]{dong2012least, yao2012least, liu2016least}. However, LSM is much more costly than RTM due to its iterative formulation \cite[]{zhang2015stable}. LSM is also fraught with practical issues including slow convergence and higher sensitivity to the migration velocity model \cite[]{zeng2017guide}. Therefore, alternative methods for mitigating migration artifacts and improving the signal-to-noise ratio are appealing.

Deep learning-based approaches have achieved state-of-the-art results for image denoising, particularly in complex, high-dimensional regimes ~\cite[]{hanumanth2020application}. Supervised deep learning techniques are one such category, which involve preparing a training dataset of noisy and clean data. Many researches have proposed variants of supervised deep learning for seismic image denoising \cite[]{zhang2019patch, liu2019poststack, zhu2019seismic, zhao2020denoising}. However, supervised learning is challenging in the context of seismic imaging, as it is difficult to create datasets that contain both noise-free labels and sufficiently span the distribution of seismic images across geologic provinces \cite[]{mosser2022comparison}. In such cases, the trained models may generalize poorly to new datasets.

Conversely, \textit{unsupervised} methods utilize the learning capability of deep neural networks without needing to assemble training datasets \cite[]{karhunen2015unsupervised}. Deep image prior (DIP) \cite[]{ulyanov2018deep} is one such workflow based on deep convolutional neural networks that is used for a variety of inverse problems associated with seismic data. At a high level, the DIP exploits the inductive bias of convolutional networks towards `natural-looking' images to regularize inverse problems (particularly those common to photographic images) such as denoising and inpainting. Since DIP is unsupervised it is an appealing approach in data-limited settings such as seismic imaging. 

DIP has been explored in a wide range of seismic data applications. For example, \cite{kong2020deep} showed that a deep image prior based on a 3D UNet could be used to interpolate missing traces in shot gathers. \cite{park2020seismic} extended this work by including a regularization term to the loss function. This regularization, which involved projection onto convex sets, improved interpolation results over large data gaps. In the context of seismic imaging, \cite{liu2020must} proposed a deep image prior for denoising pre-and post-stack seismic images, and \cite{saad2021self} proposed an attention-based UNet deep image prior with volumetric patching for denoising. 

\begin{figure*}
    \centering
    \includegraphics[width=\textwidth]{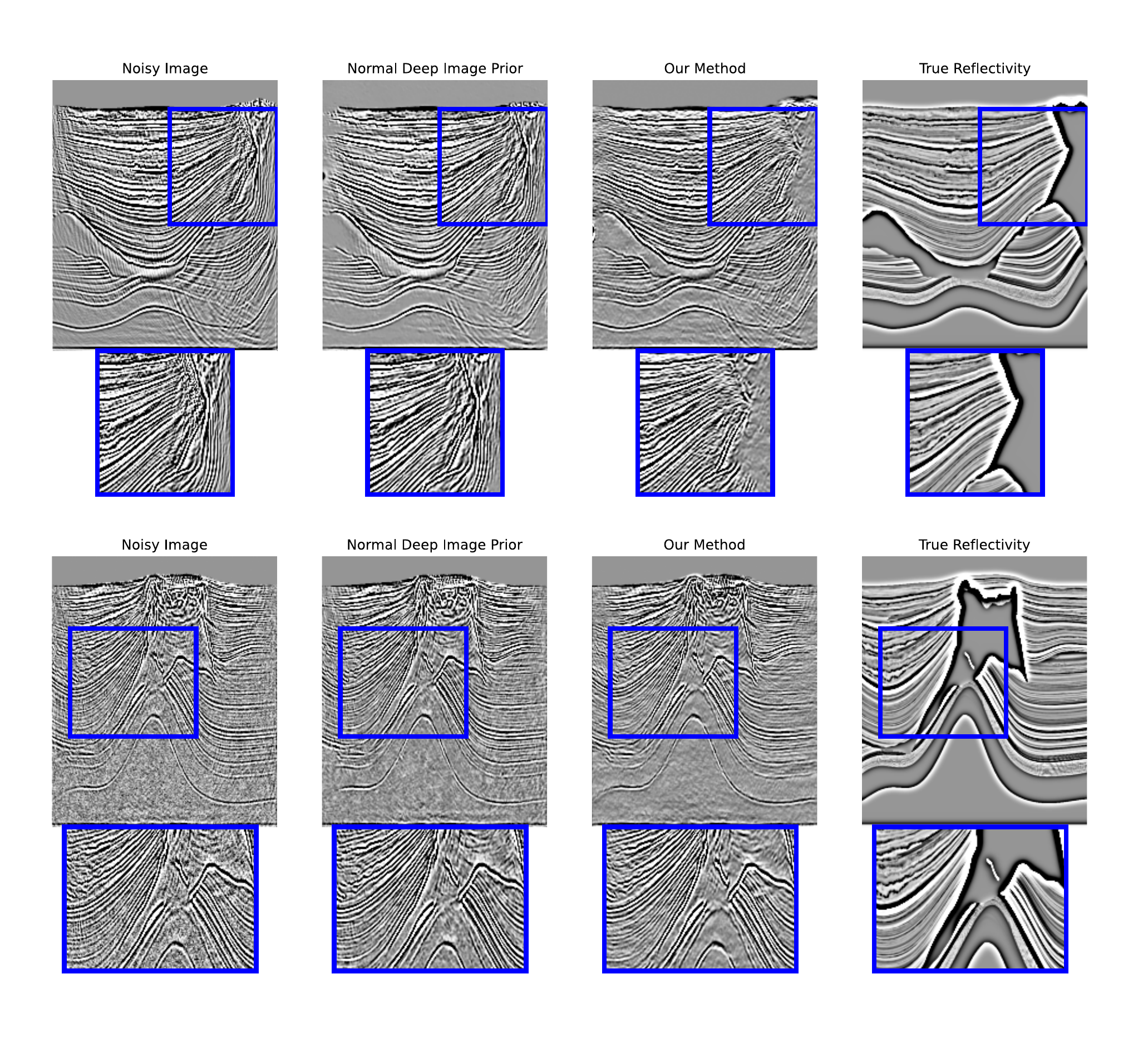}
    \caption{Top Row: Deep Image prior applied to remove migration artifacts from seismic images. By imposing a physics constraint on the reflectivity from the migration velocity, our method recovers the salt flank and base of salt better than a standard deep image prior. Furthermore, our method better attenuates migration swings in shallow sediments. Bottom Row: Deep image prior applied to removing migration artifacts \textit{and} random Gaussian noise. Our method quantitatively and qualitatively mitigates random noise better than a standard deep image prior.}
    \label{fig:overview}
\end{figure*}

We propose a modification of Deep Image Prior which is particularly suited to denoising both random and coherent artifacts in seismic migration images. Specifically, we modify the DIP to learn a pseudo-velocity model (the precise meaning of pseudo-velocity will be clarified shortly) and relate the pseudo-velocity to the RTM image using the reflectivity term of the Born system of the acoustic wave equation \cite[]{dai2013plane}. Furthermore, we introduce a regularization term to the DIP objective function which limits the magnitude of the velocity perturbation. This guides the long-wavelength features of the RTM image which helps mitigate migration artifacts that are inconsistent with the velocity model.

By incorporating these physics-based constraints in our DIP modification, we significantly improve the network's performance in removing both coherent and random noise from seismic images compared to a standard DIP implementation. Furthermore, our approach maintains the advantages of a standard application of DIP; it does not require a training dataset $\{x_i, y_i\}$ of paired noise-free examples, it does not make any assumptions about the noise distribution in the images, and it does not add significant computational expense.

\subsection{Related Work}

Other variants of unsupervised or self-supervised learning have also been proposed in the context of seismic data. For example, \cite{birnie2021potential} proposed a Noise-to-Void network \cite[]{krull2019noise2void} for denoising post-stack time-domain images. This network predicts a central pixel based on its neighboring values which helps to mitigate random noise. However, it relies on the assumption of uncorrelated noise between pixels, which may fail, for example, when considering the removal of migration parabolas.

\cite{zhao2022unsupervised} proposed an unsupervised version of a CycleGAN-based approach \cite[]{zhu2017unpaired} for denoising in the shot gather domain. They demonstrated that their network can mitigate ground roll in real data examples. \cite{wu2022self} proposed a self-adaptive architecture for denoising both coherent and random noise. However, their method assumes a mixture of Gaussian and Poisson noise distributions which may not be valid, in general. Related unsupervised learning approaches based on variants of convolutional neural network architectures have been proposed by many others \cite[]{van2021self, chen2019improving, zhang2019unsupervised, gao2021seismic}.

\begin{figure*}
    \centering
    \includegraphics[width=\textwidth]{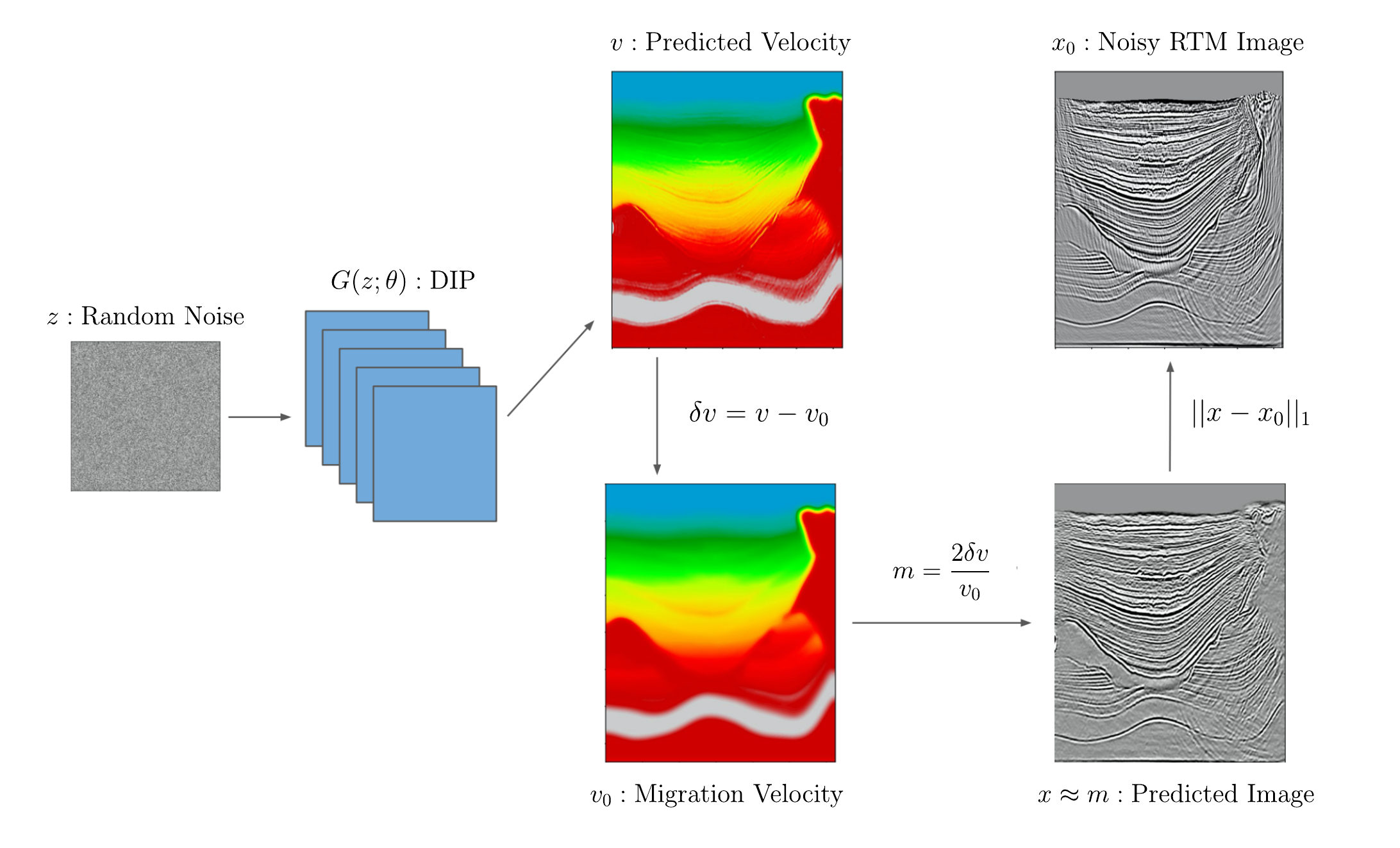}
    \caption{Schematic of our proposed workflow. The network outputs a pseudo-velocity estimate $v$, which is penalized to be close to the migration velocity $v_0$ so that the model perturbation $\delta v = v-v_0$ is small. The reflectivity, $m$, can be related to the velocity perturbation by $m = \frac{2\delta v}{v_0}$. The predicted reflectivity is then compared to a noisy RTM image $x_0$ via an L1-norm on the residual, $||x-x_0||_1$. Since we do not expect the estimated reflectivity to match the RTM image exactly, we include a weight, $w$, and pass $m$ through a single 2D convolution layer to better match the RTM image. This modifies the predicted reflectivity estimate to $x \approx m$  with $x = \textrm{conv2D}\bigg(\frac{(2+w)(G_{\theta}(z)-v_0)}{v_0}\bigg)$}.
    \label{fig:our_method}
\end{figure*}

\section{Methods}
\subsection{Deep Image Prior}

Deep Image Prior \cite{ulyanov2018deep} is an unsupervised method for solving inverse problems with deep convolutional neural networks (CNNs). The fundamental idea supporting DIP is that the structure of CNNs captures a significant portion of low-level image statistics even before optimization of the network weights. In other words, when a CNN is initialized with random weights and provided a random input image, for example, $x \in \mathbb{R}^{H\times W \times C} \sim \mathcal{U}(0,1)$, the output is significantly more `image-like' than random noise. This suggests that the network has an inductive bias towards natural-looking images. The inductive bias is attributed to the translation-invariance properties of convolutional layers and the separation of scale through pooling and unpooling layers.

To solve inverse problems with DIP, we begin with a corrupted image $x_0$ (by noise or some other operator such as a pixel mask) and a user-defined CNN architecture. We denote this architecture $G_{\theta}(z)$ where $z \in \mathbb{R}^{W \times H \times C}$ is a random input image input with distribution $z\sim \mathcal{U}(0, 1/10)$, width, $W$, height $H$, and channels $C$. The subscript $\theta$ represents the network parameters. We define a loss function that penalizes the residual between the network output and the corrupted image in some norm

\begin{equation}
\label{DIP original objective}
\mathcal{L} := ||G_{\theta}(z)- x_0||.
\end{equation}

Minimizing this loss by iterative gradient descent, while holding $z$ fixed, produces predictions $G_{\theta_j}(z) = \tilde{x}_j$ where $j$ indicates the iteration number. Before convergence of the objective function to some minimizer, clearly $\tilde{x}_j \neq x_0$. Interestingly, \cite{ulyanov2018deep} observed that the CNN will typically produce de-corrupted images before fitting corruptions in later iterations. In other words, the residual $\tilde{x}_j - x_0$ before minimizing Equation  \ref{DIP original objective} produces an estimate of the corruptions in the image. By terminating training before convergence of the loss function, a de-corrupted image can be recovered. \cite{ulyanov2018deep} provide examples of this observation in Figure 7 of their paper. 
 
In the context of inverse theory, the reconstruction process is an ill-posed inverse problem, because the solution is non-unique. In other words, many possible reconstructions are consistent with the information provided by the corrupted image \cite[]{dimakis2022deep}. According to the theory of inverse problems, the inversion can be made well posed by including regularization. Rather than imposing an explicit regularization, such as a penalty term in the objective function, the convolutional architecture of the deep image prior biases outputs towards `natural-looking' images. 

\subsection{Deep Image Prior Applied to Seismic Image Denoising}

While it has been established that a \emph{standard} Deep Image Prior\footnote{We use the term standard to indicate direct denoising of seismic images, \textit{i.e.} minimizing equation \ref{DIP original objective}.} can successfully mitigate random noise \cite[]{liu2020must, saad2021self} in seismic images, we found that it is not well suited for differentiating coherent artifacts such as migration swings from coherent reflectors in RTM images. This is exemplified in the top row of Figure \ref{fig:overview} where the salt boundary is obscured by migration artifacts when the standard DIP approach is used. 

To overcome this challenge, we propose to modify the DIP objective function in the following manner. We assume we have the noisy migration image, $x_0$, and the associated smooth background velocity, $v_0$. We then train the network to predict the velocity model, $ G_{\theta}(z)=v$ rather than the RTM image $x_0$. For simplicity, we have dropped the index $j$ indicating the optimization iterations. We use the network prediction of the velocity model to compute a velocity perturbation $\delta v=v-v_0 = G_{\theta}(z)-v_0$. We then relate $\delta v$ to the migration image through the reflectivity defined by the Born modeling system of the acoustic wave equation (see \cite{dai2013plane} for a derivation).

\begin{equation}
\label{eq:reflect}
\begin{split}
    m &= \frac{2\delta v}{v_0} \\
    x_0 &\approx m
\end{split}
\end{equation}
where $m$ is the reflectivity, $x_0$, is the RTM image, and $v$ is the true (unknown) velocity which is decomposed into a known smooth background velocity $v_0$ and a perturbation $\delta v$ such that $v = v_0 + \delta v$. Note that while $m$ represents the true amplitude reflectivity, $x_0$ is typically similar in zero-crossing location (i.e. phase) to $m$ but \textit{not} in amplitude \cite[]{dai2013plane}.

To compensate for this amplitude discrepancy, instead of relating the network output $\tilde{x}_j$ to the reflectivity by Equation \ref{eq:reflect} exactly, we introduce a modified reflectivity operator $\tilde{B}$ defined

\begin{equation}
\label{eq:born_operator}
\begin{split}
    \tilde{B}(G_{\theta}(z)) &:= \textrm{
    Conv2D}\bigg(\frac{(2+w)(G_{\theta}(z)-v_0)}{v_0}\bigg) \\ &= 
    \textrm{Conv2D}\big((2+w)m \big),
\end{split}
\end{equation}
which adds an additional weight, $w$, to the factor of two in Equation \ref{eq:reflect} and passes the estimated reflectivity image through a 2D convolutional layer with $3 \times 3$ kernel. The additional weight allows the network to learn scaling mismatches between the RTM image and the true reflectivity (since we expect them to differ), and the convolutional layer gives the network more fitting power to account for non-stationary scaling differences and other discrepancies between $m$ and $x_0$.

With these modifications, we reformulate the original DIP objective function (\ref{DIP original objective}) to

\begin{equation}
\label{eq:ourloss}
    \mathcal{L} := ||\tilde{B}(G_{\theta}(z))- x_0||_1 + \lambda||\delta v||_1, 
\end{equation}
where $\lambda$ is a regularization weight.

Relating \eqref{eq:reflect} to \eqref{eq:born_operator} shows that our network is trained to predict a pseudo-velocity. We use the term \emph{pseudo}, because the 2D convolution layer applied in \eqref{eq:born_operator} relaxes the need for $G_{\theta}(z)$ to match the true velocity exactly. The pseudo-velocity is related to the RTM image by the modified reflectivity operator---the first term of \eqref{eq:ourloss}---while penalizing large velocity perturbations $\delta v$, the second term of \eqref{eq:ourloss}. To clarify our proposed workflow, we show it schematically in Figure \ref{fig:our_method}.

Intuitively, we believe our method is advantageous because incorporating the velocity information imposes a constraint on the long-wavelength features of the reflectivity image. Because the migration artifacts are not consistent with a satisfactory velocity model, the DIP can identify and remove migration artifacts despite sharing similar wavenumber content with real reflectors.

\subsection{Data Preparation}

To test our proposed approach, we use both synthetic data with salt bodies and field data examples from the Gulf of Mexico. We build the synthetic dataset using ten 2-D slices of the SEAM Phase I Salt \textit{P}-wave velocity model. Figure \ref{fig:our_method} shows an example of this velocity model. For each slice, we prepare constant-density acoustic RTM images. For the migrations, we smooth the known velocity model with a Gaussian filter \cite[]{2020SciPy-NMeth} using an isotropic standard deviation of 8. We generate a collection of noisy images by averaging $N_k = \binom{\tau}{k}$ single-shot RTM images where $\tau=243$ is the maximum number of single-shot images. Since migration artifacts are suppressed with stacking, this allows us to assess the performance of the deep image prior as a function of the noise in the image. We evaluate our results on the noise levels $N_k \in \{122, 61, 31\}$ which corresponds to reducing the number of available single-shot images by factors of 2, 4, and 8. We maintain equal source spacing for each reduction factor. We prepare a second dataset of images with both migration artifacts and random noise by adding zero-mean Gaussian noise with variance scales of 0.05 and 0.01 to images formed by a reduction factor of $k=4$ $(N_k = 61)$ shots. This allows us to assess the ability of our method to mitigate both coherent and random noise.

The $k^{th}$-shot image is given by the zero-lag cross-correlation imaging condition

\begin{equation}
I_k(\mathbf{x}) = \nabla^2_{\mathbf{x}}\int_0^{T} u(\mathbf{x}, t)\psi(\mathbf{x}, t)\ dt.
\end{equation}

The forward wavefield $u$ satisfies the constant density acoustic wave equation

\begin{equation}
\label{eq:forward}
    \nabla^2p - \frac{1}{v_0^2(\mathbf{x})}p_{tt} = f(t, \mathbf{x}_s)
\end{equation}

where $v_0$ is the smooth velocity and $f(t, \mathbf{x}_s)$ is the source wavelet. The adjoint field $\psi(\mathbf{x}, t)$ satisfies equation \ref{eq:forward} with $f=\delta d = S(p(\textbf{x}, T-t)) - d_{obs}(\textbf{x}, T-t))$, where $S$ is a sampling operator onto the receiver locations and $d_{obs}$ represents the observed data. $d_{obs}$ is generated by solving equation \ref{eq:forward} with the true velocity model $v$. Finally, $\nabla^2_{\mathbf{x}}$ is a Laplacian filter that removes long wavelength artifacts. We also apply a geometric spreading correction proportional to $\frac{1}{z}$ (where $z$ is the depth axis) to increase illumination with depth since we do not apply illumination compensation in our RTM implementation.

To solve the constant density acoustic wave equation for the forward and time-reversed solutions involved in RTM, we use Devito \cite[]{devito-api,devito-compiler}, an open-source symbolic finite difference discretization library. Table \ref{tab:Geophysical Params} lists the relevant parameters for our simulations.

\begin{table}[H]
\caption {Geophysical Parameters Used in Preparing the Dataset}
\label{tab:Geophysical Params}
\small 
\begin{tabular}{p{0.2\linewidth}p{0.2\linewidth}p{0.5\linewidth}}
\toprule
 Parameter &  Value   & Explanation                                                 \\    \midrule
 $\Delta x$   & 40 meters   & Grid spacing in $x$ (horizontal axis) \\              
 $\Delta z$   & 20 meters   & Grid spacing in $z$ (depth axis)       \\   
 $n_x$         & 625         & Number of grid points in $x$ \\
 $n_z$         & 751         & Number of grid points in $z$ \\
 $f(t, \textbf{x}_s)$       &  Ricker Wavelet  & Wave equation forcing term   \\ 
 $f_m$        & 10 Hz   & Source wavelet peak frequency  \\
 $\tau$        & 243     & Total number of shots per slice \\
 $\Delta s$     & 100 meters & Grid spacing between shots \\
 $N_{rec}$       & 1001    & Total number of receivers for each shot \\
 $\Delta r$        & 25 meters & Grid spacing between receivers \\
 $S_z$        & 200 m & Source depth (same for all shots) \\
 $R_z$        & 200 m & Receiver depth (same for all receivers)\\
 $T$        & 12s & Integration limit in imaging condition \\
\bottomrule
\end{tabular}
\end{table}

To evaluate the performance of our proposed method with the synthetic dataset, we compare our network's output to the true reflectivity in terms of peak signal-to-noise ratio and structural similarity index \cite[]{ssim}. We benchmark our model against a standard implementation of deep image prior and simple stacking of single-shot RTM images. 

To validate our approach, we test it with field data examples from the Gulf of Mexico. The velocity model and RTM images were provided and licensed by CGG. The images formed by stacking all shots are of very high quality, owing to the high-resolution velocity model, so to test the denoising capability of our network we use only 25\% of the available shots.

\subsection{Network Architecture}

For each example, we use the `skip' architecture described by \cite{ulyanov2018deep}, a UNet-like convolutional network with an encoder-decoder structure and skip connections. The network uses five downsampling channels, five upsampling channels, and leaky-ReLU activations \cite[]{maas2013rectifier, ronneberger2015u}. For optimization, we use Adam \cite[]{kingma2014adam} and optimize for up to 8000 iterations.

\begin{figure*}
    \centering
    \includegraphics[width=0.8\textwidth]{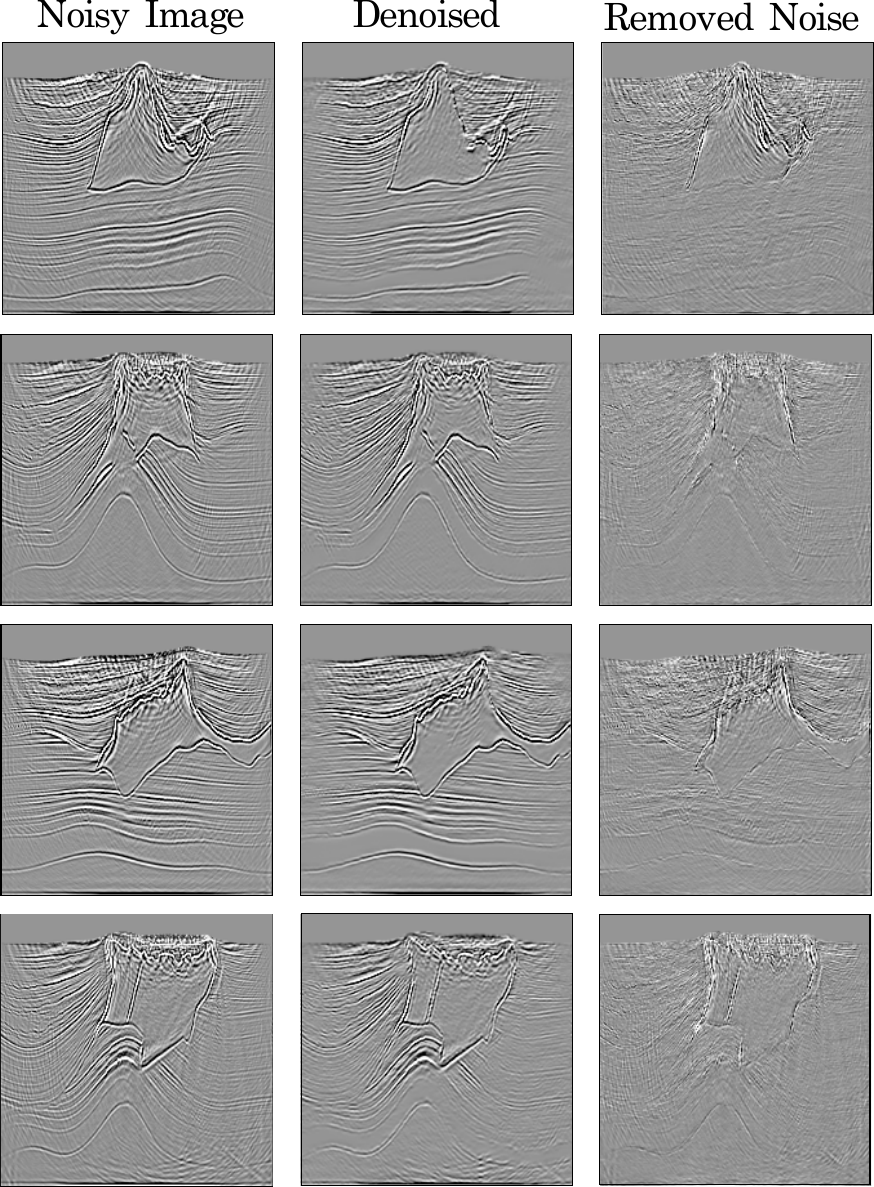}
    \caption{Denoising with the proposed Deep Image Prior modifications applied to mitigating migration artifacts in synthetic data examples. The left column shows the target image which contains migration artifacts that are particularly visible in the shallow sediments and near salt boundaries. The middle column shows our network reconstruction after 8000 iterations of network updates. The right column shows the difference between the target image and our reconstruction. Migration artifacts are mitigated, while only minor signal leakage occurs.}
    \label{fig:dip_results}
\end{figure*}

\section{Results}

Here we present the denoising results of our proposed method to the synthetic datasets described above. A quantitative evaluation of our method applied to the first dataset, which only contains migration artifacts, is shown in Table \ref{tab:numeric_results}. Quantitative evaluation of the results for the dataset with both migration artifacts and random noise are presented in Table \ref{tab:Noisy_Image_Results}.

\begin{figure*}
    \centering
    \includegraphics[width=\textwidth]{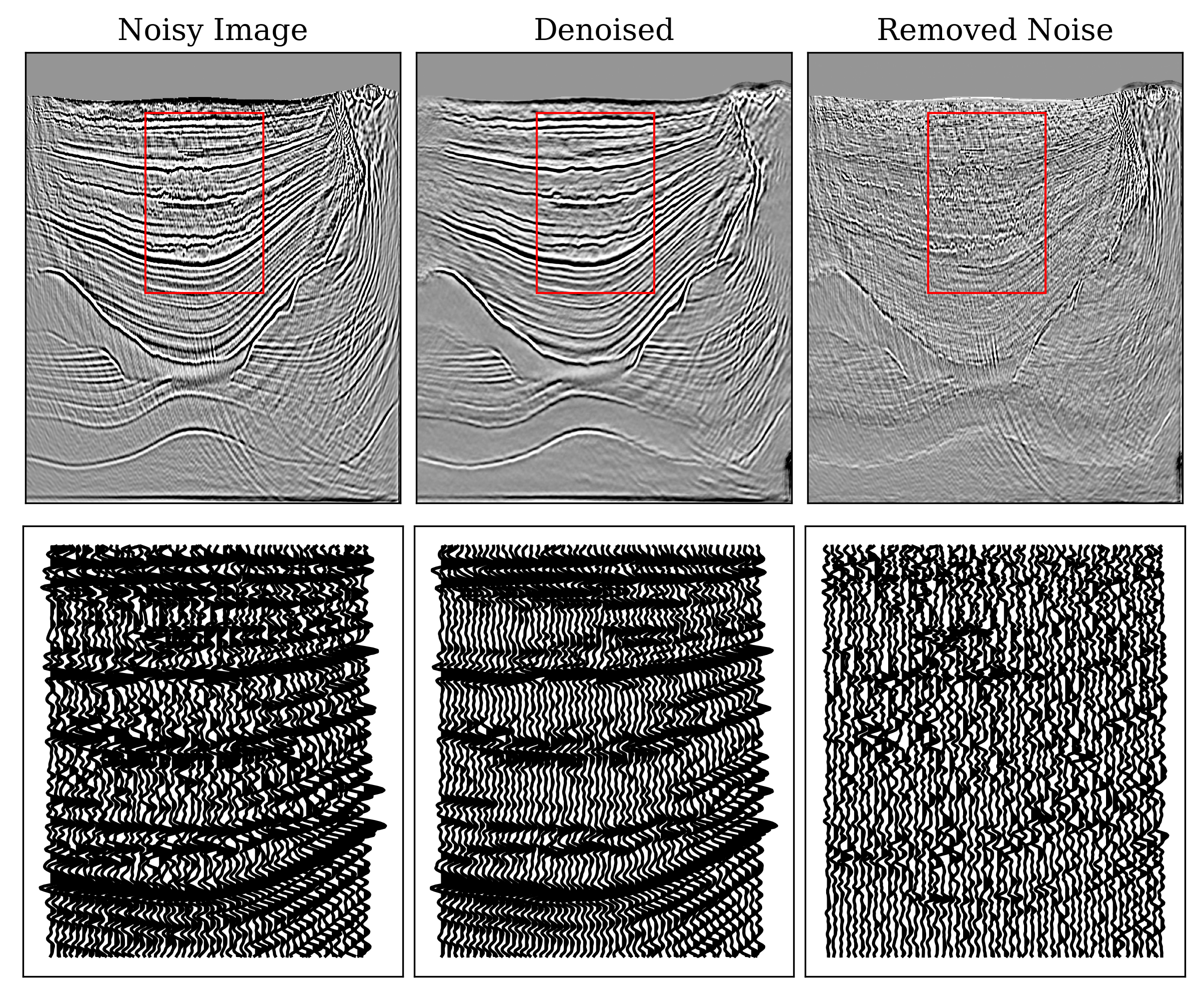}
    \caption{Denoising result emphasizing the shallow sediments showing a trace wiggle plot. Top Row: the leftmost panel shows a noisy image formed by stacking 31 shots, the middle panel shows the denoised image and the rightmost panel shows the difference between the denoised and noisy image. Bottom Row: wiggle plots of the inset region denoted by the red box.}
    \label{fig:wiggle_plots}
\end{figure*}

\begin{figure*}
    \centering
    \includegraphics[width=\textwidth]{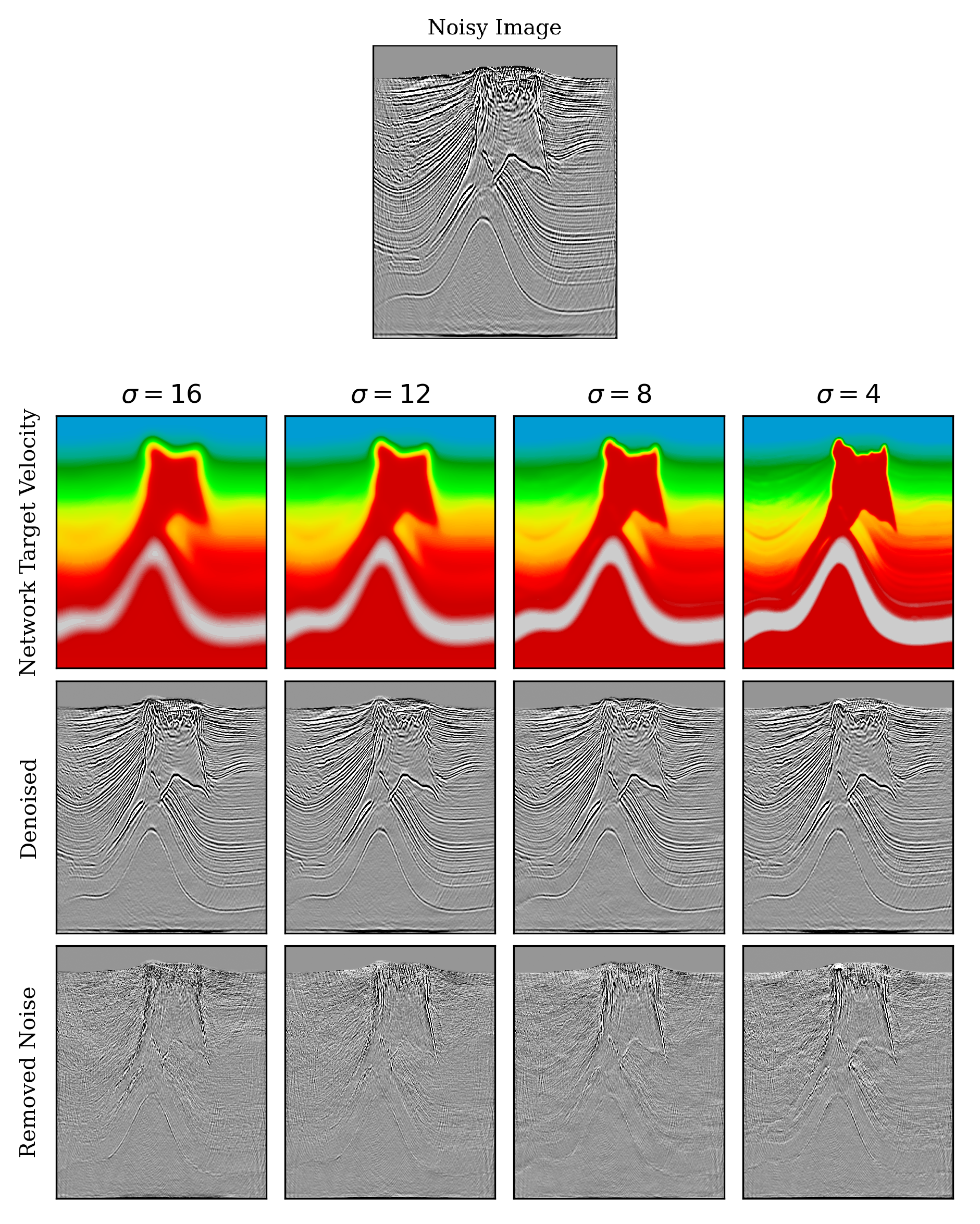}
    \caption{Test of the smoothness of the network target velocity model, $v_0$ on the denoising quality. We smooth the velocity with a Gaussian filter over four smoothness levels parameterized by the standard deviation, $\sigma$.}
    \label{fig:dip_smooth_tests}
\end{figure*}

Figure \ref{fig:dip_results} illustrates the effectiveness of our proposed method in removing migration artifacts from seismic images devoid of random noise. The left column shows the noisy target image $x_0$ which is formed by stacking single-shot migrations with a decimation factor of $k=4$ or 64 single-shot images. Migration artifacts are visible in the shallow sediments, deeper sediments, and near salt boundaries. The middle column shows the reconstruction generated by our network, while the right column shows the difference between the reconstruction and the target image. As seen in the difference plots, our method significantly mitigates the parabolic migration artifacts with only minimal signal leakage. To emphasize the improvement in the shallow sediments, Figure \ref{fig:wiggle_plots} shows a wiggle plot of the shallow sediments before denoising, after denoising, and the difference. Artifacts are significantly mitigated with minimal signal leakage. 

\begin{table}
\caption{Evaluation of denoising results for the dataset with only migration artifacts. The columns $k$ indicate the shot decimation factor. We report mean and 95\% confidence intervals for N=10 samples.}
\label{tab:numeric_results}
\centering
\begin{tabularx}{\columnwidth}{>{\hsize=.6\hsize}X>{\hsize=1.2\hsize}X>{\hsize=.6\hsize}X>{\hsize=.6\hsize}X}
\hline\hline
\multicolumn{4}{c}{Peak Signal-to-Noise Ratio}\\
 & $k=8$ (31) & $k=4$ (61) & $k=2$ (122) \\
\cline{2-4}
Stack & 18.92 $\pm$ 0.74 & 18.93 $\pm$ 0.75 & 18.94 $\pm$ 0.75 \\
Standard DIP & 18.93 $\pm$ 0.75 & 18.95 $\pm$ 0.76 & 18.95 $\pm$ 0.76 \\
Our Method & \textbf{19.02 $\pm$ 0.74} & \textbf{19.00 $\pm$ 0.74} & \textbf{19.02 $\pm$ 0.74} \\
\hline\hline
\multicolumn{4}{c}{Structural Similarity Index}\\
 & $k=8$ & $k=4$ & $k=2$ \\
\cline{2-4}
Stack & 0.48 $\pm$ 0.01 & 0.46 $\pm$ 0.02 & 0.42 $\pm$ 0.02 \\
Standard DIP & 0.50 $\pm$ 0.05 & 0.51 $\pm$ 0.05 & 0.50 $\pm$ 0.05 \\
Our Method & \textbf{0.55 $\pm$ 0.05} & \textbf{0.54 $\pm$ 0.04} & \textbf{0.55 $\pm$ 0.08} \\
\hline\hline
\end{tabularx}
\end{table}

\begin{table}[]
    \centering
    \caption{Evaluation of denoising results for the dataset with both migration artifacts and additive Gaussian noise. All samples use a decimation factor $k=4$ (61 shots). The columns $\sigma$ indicate the variance of the noise added. We report mean and 95\% confidence intervals for N=10 samples.}
\label{tab:Noisy_Image_Results}
\begin{tabular}{lcccc}
\hline\hline
\multicolumn{5}{c}{Peak Signal-to-Noise Ratio} \\
\cline{1-5}
& & & $\sigma=0.05$ & $\sigma=0.1$ \\
Stack & & & 18.79 $\pm$ 0.77 & 18.54 $\pm$ 0.84 \\
Standard DIP & & & 18.91 $\pm$ 0.75 & 18.75 $\pm$ 0.79 \\
Our Method & & & \textbf{18.98 $\pm$ 0.73} & \textbf{18.93 $\pm$ 0.76} \\
\hline\hline
& & & & \\
\multicolumn{5}{c}{Structural Similarity Index} \\
\cline{1-5}
& & & $\sigma=0.05$ & $\sigma=0.1$ \\
Stack & & & 0.42 $\pm$ 0.05 & 0.35 $\pm$ 0.06 \\
Standard DIP & & & 0.46 $\pm$ 0.05 & 0.37 $\pm$ 0.06 \\
Our Method & & & \textbf{0.55 $\pm$ 0.06} & \textbf{0.50 $\pm$ 0.05} \\
\hline\hline
\end{tabular}
\end{table}

Next, we test the effect of the background velocity's smoothness on the quality of the denoised image. To do this we progressively increase the standard deviation of a Gaussian filter on the known velocity model. These results are presented in Figure \ref{fig:dip_smooth_tests}. Even with a very smooth model, the denoising can mitigate migration artifacts.

\begin{figure*}
    \centering
    \includegraphics[width=0.7\textwidth, ]{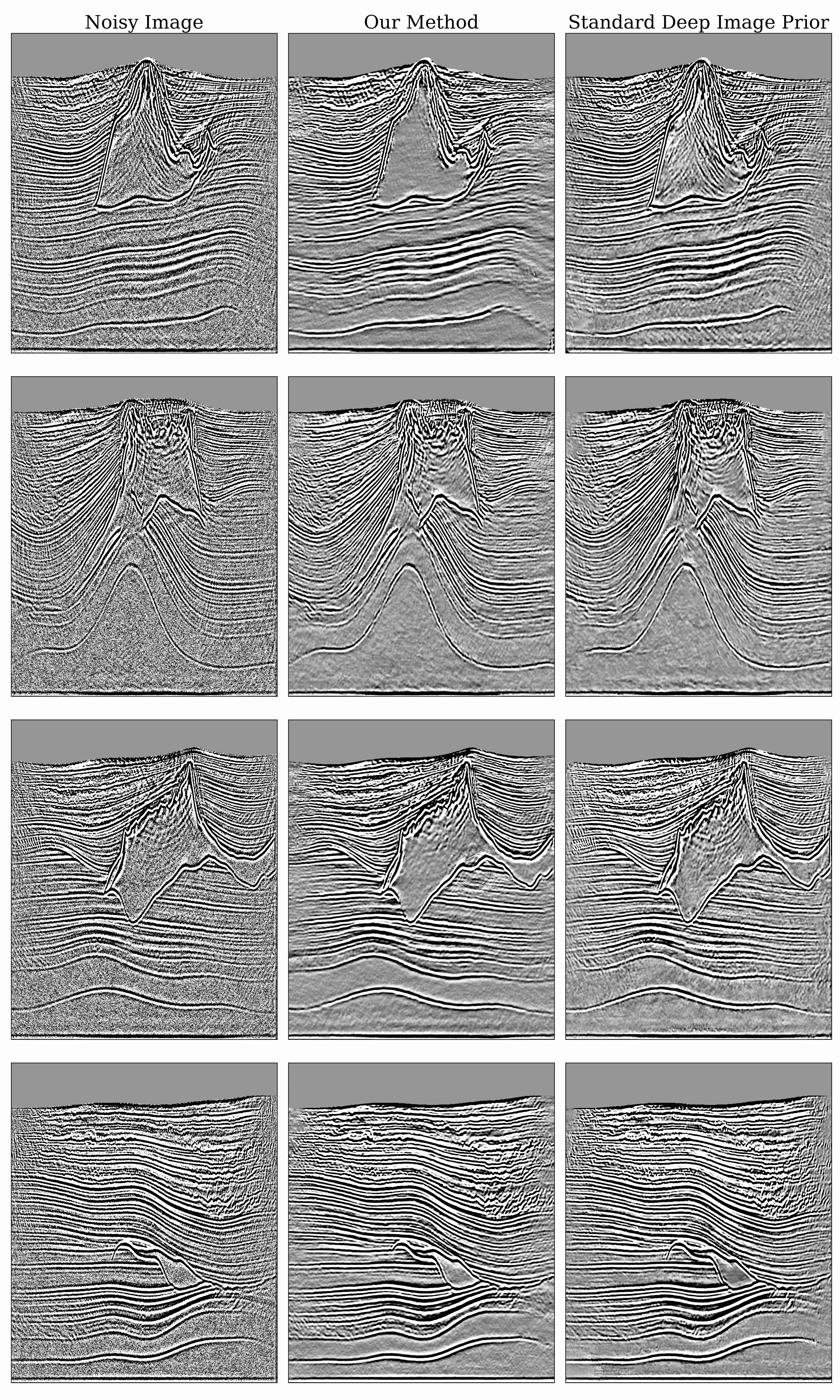}
    \caption{A comparison of our method to a standard deep image prior for mitigating both coherent migration artifacts and random noise. The left column shows a noisy image with migration artifacts and added Gaussian random noise. The middle column shows the result of our method applied to denoising the image, while the right column shows the result of a standard deep image prior.}
    \label{fig:dip_noisy_results}
\end{figure*}

We then test the efficacy of our proposed method for mitigating \textit{both} coherent migration artifacts and random Gaussian noise. Figure \ref{fig:dip_noisy_results} shows the result of our method applied to these images. The left column shows the noisy target images $x_0$ which contain migration swings and are further corrupted with additive Gaussian noise with variance $\sigma=0.05$. The images are formed by stacking single-shot migrations with a decimation factor of $k=4$ or 61 single-shot images. The middle column shows the denoising result from our proposed approach and the right column shows the denoising results from a standard deep image prior. Both our proposed approach and a standard deep image prior reasonably remove random noise. However, the salt flanks and interiors of the salt bodies are better recovered than by the standard deep image prior, we think because our method imposes constraints on the long wavelength features of the image through the velocity model. This is best observed in the first row of the figure. 

\begin{figure*}
    \centering
    \includegraphics[width=0.7\textwidth, ]{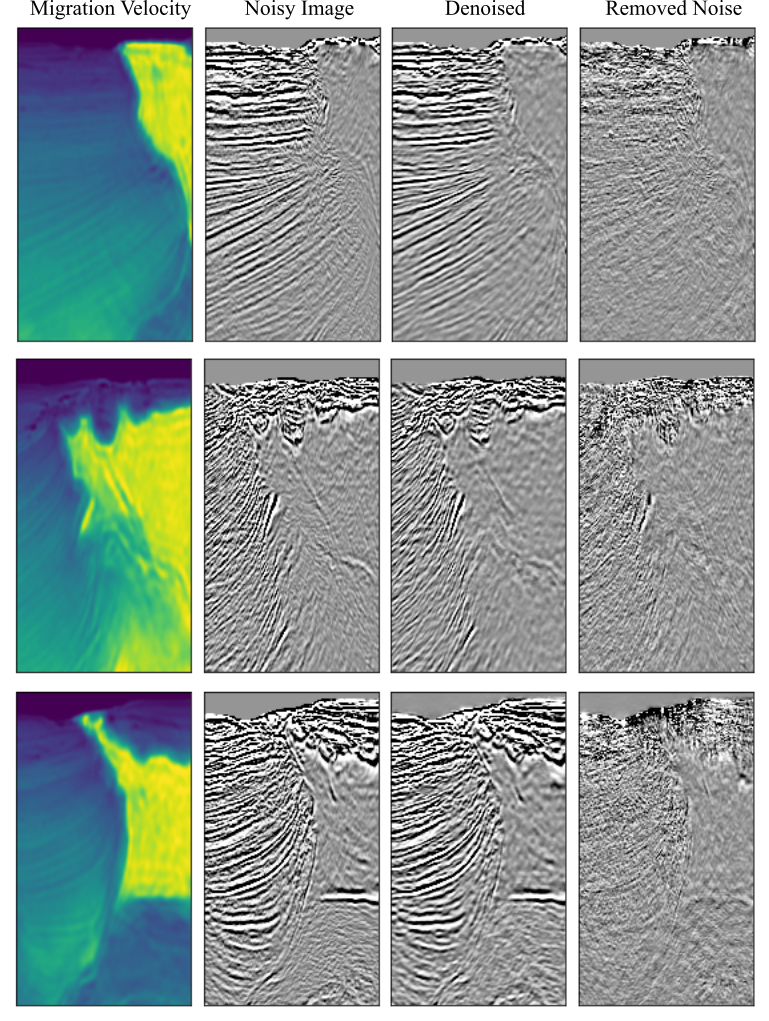}
    \caption{Application of our modified DIP denoising to field data examples from the Gulf of Mexico. The left column shows the migration velocity model associated with several field data slices (second column from left) which contains both coherent and random artifacts. The second-from-right column shows the result of our method applied to each example. The right column shows the difference between the starting image and our result. Seismic data under license from CGG.}
    \label{fig:dip-cgg-gom}
\end{figure*}

Finally, we present the result of our denoising on field data examples from the Gulf of Mexico. These are shown in Figure \ref{fig:dip-cgg-gom}. A significant quantity of noise is removed with minimal signal leakage.

\section{Discussion}

\subsection{Regularization Weight Scheduling}

The weight of the regularization term $\lambda$ has a significant impact on the resultant image quality. In particular, we encountered optimization challenges where only a single term of the objective function had a significant reduction if $\lambda$ was too large or too small. We found that applying a learning rate schedule to $\lambda$ improved the robustness of our method to this optimization challenge. We explored several strategies for tapering the value of this term including using exponential decay and linear decay as functions of the number of iterations. We found that a linear decay of the weight from $\lambda=3.0$ to $\lambda=0.4$ worked well across all examples. With this weight schedule, the generator is more heavily weighted to learn the migration velocity model in early iterations, since it emphasizes learning a small perturbation $\delta v = v - v_0$, before decreasing the penalty for larger perturbations in later iterations. This process helped guide the network towards consistent perturbations of the migration velocity model and better denoising of the RTM images.

\subsection{Early Stopping}

A primary limitation of DIP-based denoising is the potential for overfitting. The network might eventually fit both noise and signal given sufficient iterations. Therefore, the training process needs early stopping. Alternatively, checkpoints of images can be saved to disk during optimization and reviewed. Automation of the stopping criterion is an area of active research with application to both standard RGB images \cite{zhou2019towards} and seismic images \cite{zhang2023unsupervised}. In this work, however, we do not automate the early stopping process. We allow for a maximum of 8000 optimization steps, save checkpoint images every 50 steps, and choose the image checkpoint that maximizes either the signal-to-noise ratio or structural similarity. Almost always, this resulted in termination between epochs 6000 and 8000. The number of iterations will naturally depend on the learning rate, model architecture, and dataset which vary in practice. 

We found that the overfitting issues of DIP were mitigated by our loss function. We observed that several hundred to thousand additional iterations elapsed after converging to the images presented herein before overfitting to the noise in the images. However, a robust stopping criterion for our method remains to be developed.

\subsection{Amplitude Gain Control}

Because we do not apply illumination compensation in our synthetic RTM dataset, there is a very strong depth decay in amplitude in the images we produce (stronger than the expected depth decay due to increasing velocity). We found that without applying a depth-dependent gain, both the standard deep image prior and our method struggled to reconstruct deeper horizons in the images. The need for gain correction could limit the application of our method if exact amplitudes are needed. The gain correction we apply is proportional to the depth which is an invertible correction, so a reasonable amplitude recovery should be possible after applying the network correction.

\subsection{Extension to 3D}
While our method is developed and tested in 2D, it can be readily extended to be applied to a full 3D volume. However, due to GPU memory constraints, this may require distributing the network weights over several GPUs (model parallelism) for large volumes. The trend toward increasing GPU memory capacity can simplify these engineering challenges. Furthermore, memory constraints can be reduced with partially reversible architectures at the cost of additional compute time \cite[]{brugger2019partially}. 

\subsection{Computational Cost}
Our method adds very little additional computational cost compared to the standard deep image prior, adding only a handful of additional parameters to the network and floating point operations every iteration. Temporal profiling indicated that approximately 200 microseconds of wall-clock time are added per iteration relative to a standard DIP or about 1 second per 5000 thousand iterations of additional compute time. Since only 4000-8000 iterations were necessary for the results shown in this paper, this is a negligible cost. 

The absolute wall clock execution time will depend strongly on the size of the images used as the network input. For the synthetic examples on a 625 $\times$ 751 grid, the training took 2:01 minutes per thousand iterations on a single NVIDIA Tesla V100 GPU. The total training time for a single slice is thus between 8 and 16 minutes. Reconstructing a 3D volume from 2D slices would be an embarrassingly parallel process, adding little additional overhead to denoising an entire volume. However, we speculate that a fully 3D approach would produce more consistent results across slices.

\section{Conclusion}

Our modified Deep Image Prior approach, incorporating velocity-model-based constraints, effectively reduces both coherent and random noise in seismic images. This method enhances image quality without requiring extensive training datasets, maintaining computational efficiency. Future work will explore extending this approach to 3D seismic data and refining the early stopping criteria to improve performance and applicability further.






\bibliographystyle{abbrvnat}
\bibliography{reference}



\end{document}